\documentclass[aps,pra,twocolumn,floatfix,10pt]{revtex4-1}

\usepackage{amsfonts,amsmath,amssymb,amsthm,amscd}
\usepackage{acronym,array,bm,color,dcolumn}
\usepackage{graphicx,nomencl,revsymb4-1,upgreek,url}
\usepackage{hyperref}
\hypersetup{colorlinks=true, pdfauthor=Uzoma Onunkwo, pdftitle=Practical AQC, citecolor=cyan, urlcolor=blue}

\newcommand{\ket}[1]{\ensuremath{\left\vert{#1}\right\rangle}}

\newcommand{\order}[1]{\mathcal{O}\left( #1 \right)}

\newcommand{\cbb}[1]{\ensuremath{[\![{#1}]\!]}} 

\definecolor{darkgreen}{rgb}{0,0.4,0}
\definecolor{purple}{rgb}{0.4,0,0.8}
\definecolor{cyan}{rgb}{0,0.8,0.9}

\begin{document}

\title{A family of $\cbb{6k,2k,2}$ codes for practical, scalable adiabatic quantum computation}

\author{Anand Ganti}
\email[Electronic address: ]{aganti@sandia.gov}
\affiliation{Department of Advanced Information \& Networking Systems Engineering (09336),
Sandia National Laboratories, Albuquerque, NM 87185, USA}
\author{Uzoma Onunkwo}
\email[Electronic address: ]{uonunkw@sandia.gov}
\affiliation{Department of Advanced Information \& Networking Systems Engineering (09336),
Sandia National Laboratories, Albuquerque, NM 87185, USA}
\author{Kevin Young}
\email[Electronic address: ]{kyoung@sandia.gov}
\affiliation{Department of Scalable and Secure Systems Research (08961),
Sandia National Laboratories, Livermore, CA 94551, USA}

\date{\today}

\begin{abstract}
\noindent
In this work, we introduce a new family of $\cbb{6k,2k,2}$ codes designed specifically to be compatible with adiabatic quantum computation.  These codes support computationally universal sets of weight-two logical operators and are particularly well-suited for implementing dynamical decoupling error suppression.  For Hamiltonians embeddable on a planar graph of fixed degree,  our encoding maintains a planar connectivity graph and increases the graph degree by only two.  These codes are the first known to possess these features.  
\end{abstract}

\maketitle

\section{Introduction}
\label{sec:intro}
\noindent Adiabatic quantum computation (AQC) has garnered considerable attention from the research community as a possible alternative to the resource-intensive circuit model.  While provably equivalent to other complete models of quantum computation~\cite{aharonov2008adiabatic, mizel2007simple}, AQC further exhibits a number of unique properties, such as a gap-protected ground state and inherent robustness to dephasing.  These features have driven many to suspect that AQC may require far fewer physical resources to implement than other paradigms of quantum computing (QC).  Unfortunately, it is not yet known how one might leverage these features to yield a fault-tolerant model of computation, and the search for an adiabatic threshold theorem remains the major outstanding theoretical goal of the field \cite{Lidar2007DDAQC}.

Recall that in AQC, a set of qubits are first prepared in the ground state of an applied initial Hamiltonian $H_0$.  Typically, $H_0$ has a fixed energy gap between the unique ground state and the first excited state(s), and its ground state is a product state, making it easy to prepare via various means such as cooling or the use of a simple quantum circuit.  This system Hamiltonian is then slowly evolved to a final Hamiltonian, $H_1$, designed so that its ground state encodes the solution to a problem of interest.  The adiabatic theorem guarantees that, so long as the Hamiltonian is changed sufficiently slowly, the state of the system will track the ground state of the applied Hamiltonian.  The final state of the system will therefore be the ground state of $H_1$, and measurement of this state will reveal the solution to the problem \cite{farhi2000}.

Because the system is expected to always be in its ground state, AQC is robust to many of the decoherence models which plague circuit model quantum computation, such as relaxation and dephasing in the energy eigenbasis.  However, an adiabatic quantum computer would likely be subject to a number of noise sources capable of causing excitations from the adiabatically evolving ground state and thereby corrupting the computation.  Dealing with such errors in circuit model QC necessitates the use of high-distance quantum codes, but such codes must possess high-weight logical and stabilizer operators that cannot be easily accommodated within the Hamiltonian formalism of AQC~\cite{young2013,sarovar2013}.

Low distance codes, however, have been shown capable of \emph{suppressing} errors in AQC.  The pioneering error suppression approach, known as energy gap protection (EGP), was developed by Jordan \emph{et al.}~\cite{jordan2006error} and involves encoding the AQC in a stabilizer code and adding the code's stabilizer generators to the system Hamiltonian.  This addition does not disturb the computation but introduces an energy gap which penalizes the states in the error subspace of the code.  If the noise couples to the system through  error operators detectable by the code, and the power spectrum of the noise is decreasing, then this energy penalty can dramatically reduce the transition rate, as implied by Fermi's Golden Rule.  An alternate approach, developed in~\cite{Lidar2007DDAQC}, also requires encoding the AQC in a stabilizer code but applies the stabilizer generators as dynamical decoupling (DD) pulses.  Since the stabilizers commute with the system Hamiltonian, their application as DD pulses does not affect the computation.  It was shown in~\cite{young2013,sarovar2013} that EGP and dynamical decoupling for AQC (DDAQC) can be unified under the same dynamical framework.

While both DD and EGP are promising tools for reducing error rates in AQC, realizing these approaches in a practical hardware setting requires additional resources beyond those of standard AQC -- the implementation of precise unitary control operators for DD requires exactly the resources AQC was constructed to avoid, while the inclusion of high-weight operators in the encoded Hamiltonian for EGP can only be accomplished with gadget perturbation techniques~\cite{jordan2008perturbative}.  Approximate implementation of DDAQC is likely to be more tractable since even high-weight stabilizers, when applied as sequences of unitary pulses, can be approximately decomposed as product operators on individual qubits.  Lidar~\cite{Lidar2007DDAQC} showed that one can use the $\cbb{2k+2,2k,2}$ CSS code constructed by Gottesman~\cite{gottesman1997stabilizer} to implement DDAQC using just a weight-two system Hamiltonian, see Fig.~\ref{fig:2kp22k2}.  This particular scheme, however, is unlikely to scale, as the ``hub-and-spoke'' connectivity of the code requires the two hub qubits to interact with all of the $2k$ spoke qubits.  Such high-degree of interaction is unlikely to be achieved in a physical device.

To overcome these shortcomings, we present a new family of $\cbb{6k,2k,2}$ quantum error detecting codes which maintain constant degree and weight-two interactions, thus enabling practical implementation of AQC with quantum error suppression.  We begin our discussion in Section \ref{sec:requirements} by considering the unique demands placed on quantum codes by the Hamiltonian formulation of AQC.  We then introduce the $\cbb{6k,2k,2}$, providing efficient initialization schemes, demonstrating its utility for quantum error suppression, and show that it can be implemented in a wide variety of experimental systems.  
 
\section{Desirable properties of a quantum code for DDAQC}
\label{sec:requirements}
\noindent Many interesting problems in physics and optimization may be solved by determining the ground state of some 2-local Hamiltonian, which we'll refer to as the \emph{problem Hamiltonian}.  The quantum adiabatic algorithm was designed to be a (possibly) efficient way to produce these ground states by exploiting adiabatic eigenstate dragging.  Unfortunately, and as discussed earlier, any physical realization of this algorithm will be subject to noise which may cause excitations from the adiabatically evolving ground state.  Left unchecked, these excitations will spoil a computation.  To maintain a useful device, this noise must be protected against by using some type of quantum coding, which distributes quantum information across a number of qubits.  Unless great care is taken with the choice of code, however, the encoded Hamiltonian may be extraordinarily difficult, or even impossible, to implement in hardware.  Assuming that the original problem Hamiltonian \emph{can} be easily implemented, we seek a quantum coding scheme which preserves the following properties:
\begin{enumerate}
	\item \emph{Weight-two operators}: Quantum codes identify weight-one operators as errors, while physics generally
	 forbids operators of weight-three or higher.  We therefore desire a code admitting only weight-2 logical
	 operators for all interactions. 
	 In fact, it is further desirable that the interactions be limited in type, for example $\{XX,ZZ\}$
	 or $\{XZ, ZX\}$.  These operator sets, in fact, have been shown to be universal for
	 AQC~\cite{PhysRevA.78.012352}, so this property is not overly restrictive.

	\item \emph{Fixed degree}: The AQC Hamiltonian should be such that each qubit is 
	involved in only a fixed number of two-qubit operators independent of the system size, \emph{i.e.}, the interaction graph is of bounded degree.  
	Maintaining a growing number of interactions on a single qubit will require unreasonably 
	demanding control resources and is hence deemed impractical.

	\item \emph{Planarity}: The above two properties imply that the many body component of the 
	AQC Hamiltonian can be mapped to a fixed degree graph.  However, we also desire that the AQC 
	be realizable on a \emph{planar} hardware graph, simply because such graphs are dramatically easier to implement  
		\footnote{Note that some physical architectures permit, or even demand, 
		that this property be relaxed, as illustrated by the \emph{Chimera} graph 
		used in the D-Wave architecture~\cite{choi2010embdng}.}.
	A planar problem Hamiltonian should therefore remain planar upon encoding.
\end{enumerate}

\begin{figure}
\centering
\includegraphics[width=0.8\columnwidth]{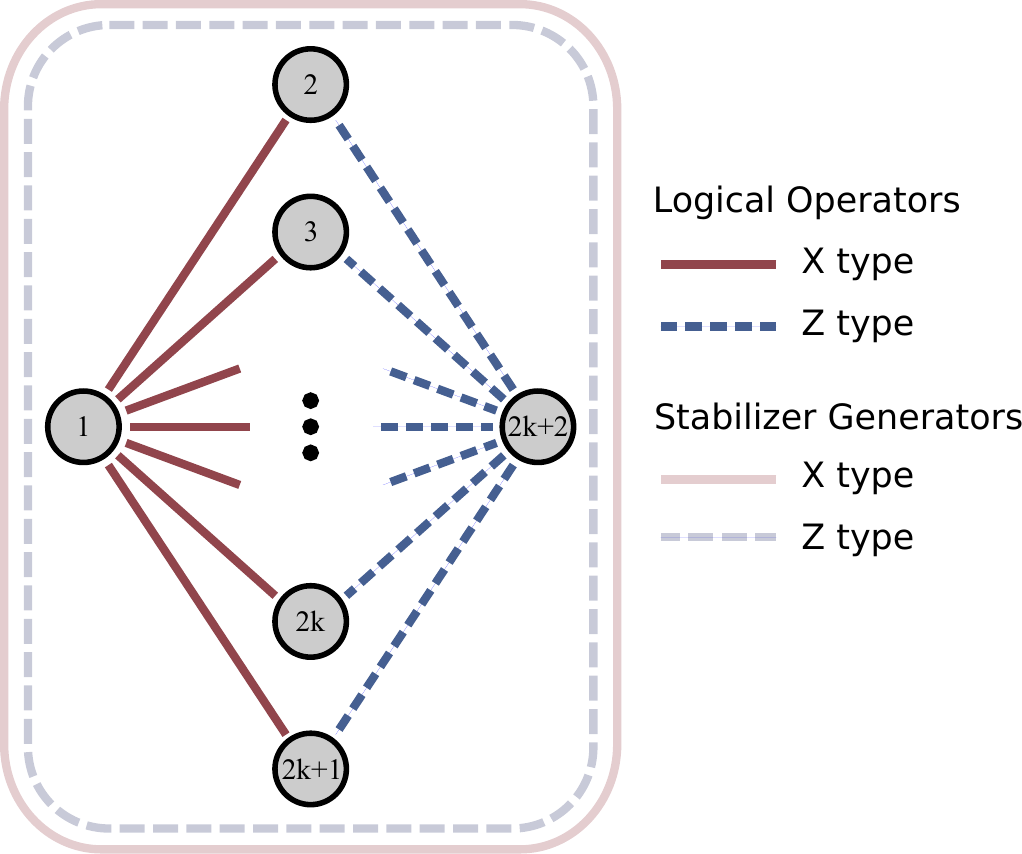}
\caption{(Color online) The Gottesman $[[2k+2,2k,2]]$ quantum error detecting code.  Notice the hub-and-spoke configuration taken by the single-qubit logical operators. Solid lines indicating $\overline{X}=XX$ operators and dashed lines indicating $\overline{Z}=ZZ$ operators.  The stabilizer generators are the $(2k+2)$-body operators, $X^{\otimes 2k+2}$ and $Z^{\otimes 2k+2}$. The high connectivity demanded of physical qubits $1$ and $2k+2$ makes this code challenging to implement for $k$ much larger than $2$.}
\label{fig:2kp22k2}
\end{figure}

As an example of a well-known code which does \emph{not} preserve all of these properties, consider the Gottesman $\cbb{2k+2, 2k, 2}$ code, illustrated in Fig.~\ref{fig:2kp22k2} 
and defined in terms of its two $(2k+2)$-body stabilizer generators, 
\begin{equation}
\label{eq:2kp22k2stab}
	S = \left\langle \bigotimes_{i=1}^{2k+2} X_i,\;\;\bigotimes_{i=1}^{2k+2} Z_i \right\rangle,
\end{equation}
and its logical operators,
\begin{align}
	\overline{X_i} 		&= X_1X_{i+1}\notag\\
	\overline{Z_i} 		&= Z_{i+1}Z_{2k+2}\notag\\
	\overline{X_iX_j} 	&= X_{i+1}X_{j+1}\notag\\
	\overline{Z_iZ_j} 	&= Z_{i+1}Z_{j+1}\label{eq:2kp22k2log}
\end{align}
where $i \in \{1, 2, \ldots, 2k\}$.
This universal set of logical operators is composed entirely of weight-two physical operators, so this code nicely satisfies property (1) above.  However, this code uses the first (last) physical qubit in all $2k$ of the logical $\overline{X}$ ($\overline{Z}$) operators and so does not satisfy property (2).  As this code grows, the coupling demands on the first and last qubits quickly become physically untenable.  Finally, notice that this code yields a planar Hamiltonian only when the unencoded Hamiltonian has linear, nearest neighbor couplings.   

\section{The \texorpdfstring{$\cbb{6k,2k,2}$}{[[6k,2k,2]]} quantum CSS code}
\label{sec:6k2k2description}
\begin{figure}[t]
\centering
\includegraphics[width=0.85\columnwidth]{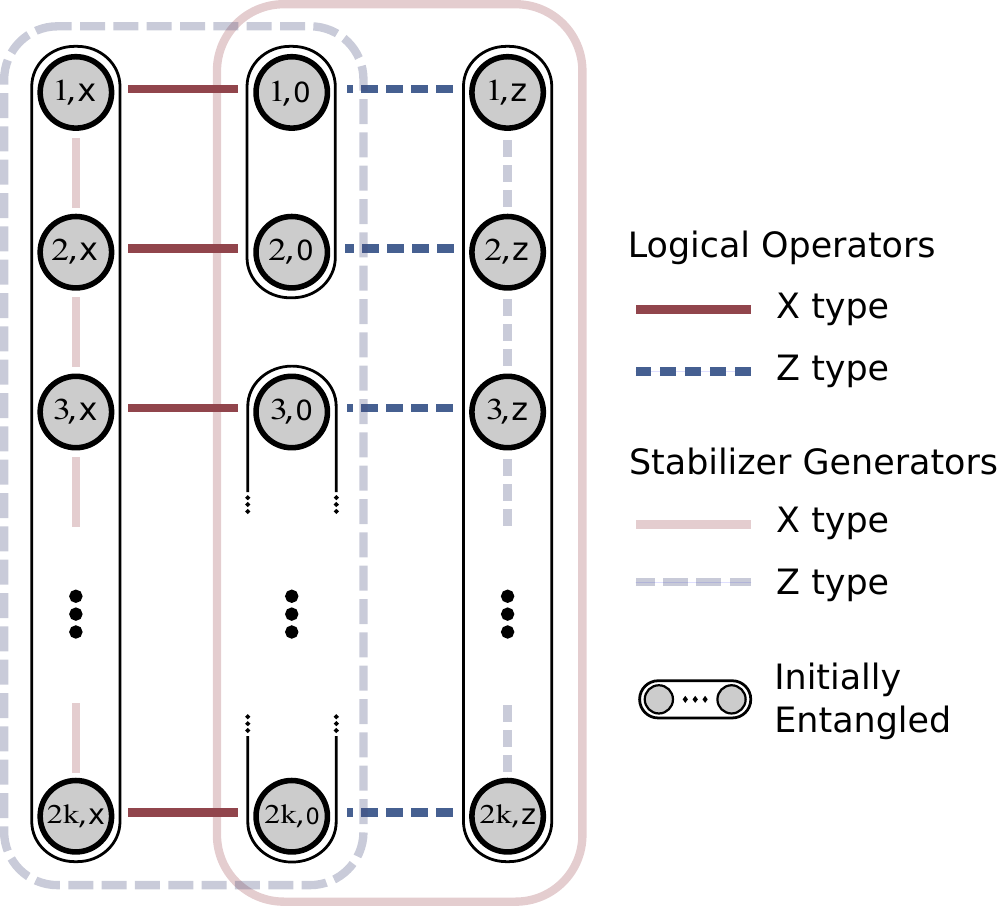}
\caption{(Color online) Schematic for a linear layout of the $\cbb{6k,2k,2}$ code indicating the physical qubits involved in each of the stabilizer generators (pale lines) and single-qubit logical operators (dark lines).  Each logical qubit is composed of a row of three physical qubits.  All code operators are weight-two with the exception of two weight-$4k$ stabilizer generators.  Qubits which are initially entangled are enclosed in black ovals.}
\label{fig:6k2k2}
\end{figure}

\begin{figure}
\includegraphics[width=\columnwidth]{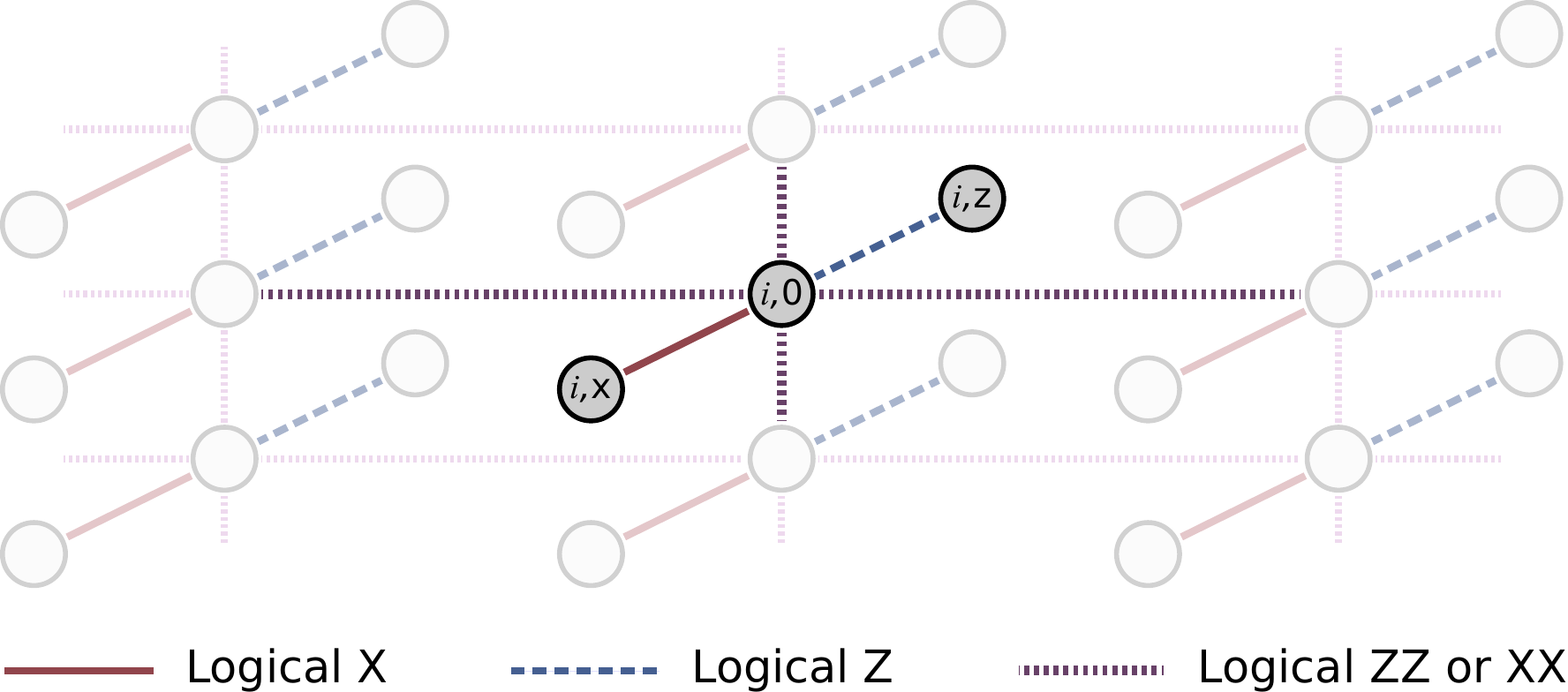}
\caption{(Color online) A 2D grid of logical qubits remains planar upon encoding in the $\cbb{6k,2k,2}$ code. We have emphasized a single logical qubit and its interactions.  Again, all interactions between logical qubits take place through the $(i,0)$ qubits.}
\label{fig:6k2k2grid}
\end{figure}

\noindent In this section, we introduce a family of $\cbb{6k,2k,2}$ quantum CSS codes that \emph{do} preserve the properties listed above.  Our code replaces each logical qubit, labeled by index $i$, with three physical qubits, labeled by the ordered pairs, $\{(i,{\rm x}),(i,{\rm 0}),(i,{\rm z})\}$, as illustrated in Fig.~\ref{fig:6k2k2}.  
The stabilizer group is generated by $4k$ operators: 
\begin{align}
\label{eq:6kstabs}
	S = \Big\langle 
		&\{X_{(i,{\rm x})}X_{(i+1,{\rm x})}\}_{i=1}^{2k-1},	\;\bigotimes_{i=1}^{2k} X_{(i,{\rm 0})}X_{(i,{\rm z})}, \notag
	\\	&\{Z_{(i,{\rm z})}Z_{(i+1,{\rm z})}\}_{i=1}^{2k-1}, \;\bigotimes_{i=1}^{2k} Z_{(i,{\rm x})}Z_{(i,{\rm 0})} \Big\rangle,
\end{align}
where $X_{(i,{\rm z})}$ is a Pauli $X$ operator acting on physical qubit $(i,{\rm z})$.  Note that it is the requirement that the two many-body stabilizer generators commute that forces the total number of logical qubits in the code to be even.  The single-body logical operators of the code are given by:
\begin{align}
	\overline{X_i} &= X_{(i,{\rm x})}X_{(i,{\rm 0})} \notag\\
	\overline{Z_i} &= Z_{(i,{\rm 0})}Z_{(i,{\rm z})}.
\end{align}
Two-body logical operators may be constructed by multiplying the associated single-body logical operators, though this at first would seem to result in a four-body physical operator, \emph{i.e.}, $\overline{X_iX_j} = X_{(i,{\rm x})}X_{(i,{\rm 0})}X_{(j,{\rm x})}X_{(j,{\rm 0})}$.  However, we may exploit the fact that logical operators are \emph{equivalent up to multiplication by elements of the stabilizer group}.  That is, any logical operator $\overline{L}$ acts identically to $s \overline{L}$ on states in the codespace, for some $s$ in the stabilizer group.  In the case of the $\cbb{6k,2k,2}$ code, all two of the form $X_{(i,{\rm x})}X_{(j,{\rm x})}$ are in the stabilizer group.  This observation allows us to write the two-body logical operators as
\begin{align}
	\overline{X_i X_j} &= X_{(i,{\rm 0})}X_{(j,{\rm 0})} \notag\\
	\overline{Z_i Z_j} &= Z_{(i,{\rm 0})}Z_{(j,{\rm 0})},
\end{align}
where we have used a similar line of reasoning in the construction of the Z-type operators.  Notice that all single body interactions are limited to the three qubits comprising the logical qubit, while all couplings involve only the $(i,{\rm 0})$ qubits.  This code therefore preserves the connectivity of the problem Hamiltonian (so a planar Hamiltonian remains planar upon encoding), and only increases the degree of the connection graph by two, as shown in Fig.~\ref{fig:6k2k2grid}.

We may establish the distance of this code by noting that all single-physical-qubit Pauli errors anticommute with at least one of the stabilizer generators, and are therefore detectable by the code. However, there are a number of single qubit errors (such as $X_{(i,{\rm 0)}}$ and $X_{(j,{\rm 0})}$) which produce an identical error syndrome, implying that the code is not capable of error \emph{correction}.  This further implies the existence of two-qubit errors (such as the two-body logical operator, $X_{(i,{\rm 0})}X_{(j,{\rm 0})}$) which do not anticommute with \emph{any} of the stabilizer generators and are therefore undetectable.  The code distance is therefore fixed at $d=2$.

\subsection*{Initialization of the encoded computation}
\label{subsec:init}
\noindent For the code to enable adiabatic quantum computation with error suppression, it must permit the easy initialization of the system into a logical state which (i) is in the codespace of the $\cbb{6k,2k,2}$ code and (ii) is the ground state of a simple, spatially local logical (encoded) Hamiltonian.  A state which satisfies these requirements is given by 
\begin{enumerate}
	\item All qubits $(i,{\rm x})$ are in the entangled state
		\begin{equation*}
			\frac{1}{\sqrt{2}} \left(\ket{+++ \cdots} + \ket{--- \cdots}\right)
		\end{equation*}
	\item All qubits $(i,{\rm z})$ are in the entangled state
		\begin{equation*}
			\frac{1}{\sqrt{2}} \left(\ket{000 \cdots} + \ket{111 \cdots}\right)
		\end{equation*}
	\item All qubit pairs $(2i\!-\!1,{\rm 0})$ and $(2i,{\rm 0})$ belong to a Bell pair,
		\begin{equation*}
			\frac{1}{\sqrt{2}} \left(\ket{00} + \ket{11}\right) = \frac{1}{\sqrt{2}} \left(\ket{++} + \ket{--}\right)
		\end{equation*}
\end{enumerate}
One may verify that this state is unchanged (\emph{stabilized}) by the stabilizer generators of the code.  Furthermore, it is stabilized by the following set of logical operators:
\begin{align}
 \overline{Z_1Z_2}, \,\,\overline{Z_3Z_4}, \,\,\ldots, \,\,\overline{Z_{2k-1}Z_{2k}} \notag \\
 \overline{X_1X_2}, \,\,\overline{X_3X_4}, \,\,\ldots, \,\,\overline{X_{2k-1}X_{2k}} \notag
\end{align}
Together with the stabilizer generators, these logical operators form a complete set of $6k$ commuting observables on $6k$ qubits.  Our initial state is therefore the \emph{unique} state stabilized by these operators and the \emph{only} state in the codespace which is the ground state of the following Hamiltonian:
\begin{equation}
\label{eq:ham_ground}
	H = -\sum_{i=1}^k \left(Z_{(2i\!-\!1,{\rm 0})}Z_{{(2i,{\rm 0})}} + X_{{(2i\!-\!1,{\rm 0})}}X_{{(2i,{\rm 0})}}\right),
\end{equation}
which will therefore assume the role of the initial Hamiltonian in the adiabatic algorithm.  
 
The standard initial Hamiltonian in AQC is a sum of Pauli X operators on each qubit.  In that case, one may simply wait for the qubits to relax to the ground state. For our code, however, we are hoping to prepare a complicated entangled state for which the thermalization time may be expected to be unreasonably long.   Fortunately, the adiabatic algorithm itself may be used to efficiently construct these states.  

Consider the large ``cat'' state occupied by the $(i,{\rm x})$ qubits.  This state may, in fact, be prepared by an adiabatic interpolation that begins with  $(i,{\rm x})$ qubits in the unique ground state of $H_0 = -\sum_{i=1}^{2k} X_{(i,{\rm x})}$ and ends in the ground space of the Ising Hamiltonian, $H_1 = -\sum_{i=1}^{2k-1}Z_{(i,{\rm x})}Z_{(i+1, {\rm x})}$.  The ground space of the Ising model is degenerate, however, so we must ensure that the system ends up in the desired cat state.  Fortunately, this happens automatically.  The adiabatic interpolation begins with the system in a $+1$ eigenstate of the operator $\bigotimes_{i=1}^{2k}X_{(i,{\rm x})}$.  This operator commutes with the Hamiltonian at all points in the interpolation, and so its expectation value will not change over the course of the adiabatic evolution.  The final state will therefore be the symmetric cat state,
\begin{equation*}
	\frac{1}{\sqrt{2}} \left(\ket{000\cdots}+\ket{111\cdots}\right),
\end{equation*}
as desired.  The $(i,{\rm z})$ cat state may be prepared similarly by replacing $X\leftrightarrow Z$ in the above procedure, as may the Bell pair states occupied by the $(i,{\rm 0})$ qubits.

\subsection*{Practical error suppression in AQC using the \texorpdfstring{$\cbb{6k,2k,2}$}{[[6k,2k,2]]} code}
\label{sec:errorsuppress}
\noindent The $\cbb{6k,2k,2}$ family of codes has a distance of $d=2$ and, as such, the best one can hope for is to improve the encoded system's robustness to weight-one noise.  Fortunately, nearly all environmental noise models couple into the system through weight-one operators, and these couplings may then be suppressed by the EGP or DD techniques discussed earlier.  The stabilizer group for the code contains the operators $X_{\rm all}=X^{\otimes 6k}$ and $Z_{\rm all}=Z^{\otimes 6k}$. These operators generate the universal decoupling group~\cite{lidar2012review}, and so may be applied in sequence to implement DDAQC.  

The EGP approach is hindered by the fact that two of the stabilizer generators are weight-$4k$ and so cannot be implemented as Hamiltonian penalty terms.  However, the results of \cite{young2013,sarovar2013} indicate nearly equivalent performance of DD and EGP.  This implies that a hybrid approach may be useful, whereupon the weight-two stabilizer generators are included as Hamiltonian penalty terms, while the weight-$4k$ stabilizer generators are applied as unitary operators to implement DD.  

Such a hybrid approach is particularly advantageous because it minimizes the errors introduced when approximating many-body unitary operators as product operators. Many-body DD pulses can be implemented several ways, including: (i) by applying the many-body operator as a Hamiltonian, for example,
\begin{align}
	XXX\!\ldots 	
		&= \exp\left(-i\frac{\pi}{2}\left(XXX\ldots\right)\right), \notag
\end{align}
or (ii) by applying a Hamiltonian composed of spatially local operators, for example,
\begin{align}
	XXX\ldots 	
	 	&= \exp\left(-i\frac{\pi}{2}\left(XII\ldots + IXI\ldots + \cdots \right)\right). 
\end{align}
Though seemingly the most straightforward, method (i) is impractical, as the many-body operator is unlikely to be directly applicable to the system.  Method (ii), on the other hand, has the advantage of easy implementation.  However, these decoupling pulses will be applied in the presence of the adiabatic Hamiltonian, which commutes with the many-body operator, but not with the spatially local operators.  Method (ii) therefore introduces a small error each time a pulse is applied.  We can quantify the magnitude of this error by assuming that the control Hamiltonian implementing the DD, $H_{\rm C}$, operates on timescales fast compared to changes in the encoded adiabatic Hamiltonian, $\overline{H_{\rm AQC}}$. A Zassenhaus/Taylor expansion \cite{Suzuki1997} of the resulting unitary operator then gives
\begin{align}
U(\tau)&= \exp \left(-i \tau \left(\overline{H_{\rm AQC}}+ \alpha H_{\rm C}\right) \right) \notag\\
&\simeq\exp(-i\overline{H_{\rm AQC}} \tau) U_{\rm DD} \left(1+ \frac{\tau \pi}{4}[\overline{H_{\rm AQC}},H_{\rm C}] \right) \notag
\end{align}
where $\exp(-i\alpha H_{\rm C}\tau)= U_{\rm DD}$ is the DD pulse, $\alpha \tau = \pi/2$, and the expansions have been taken to order 
$\tau^2$. The norm of the lowest-order error term can be bounded as $||[\overline{H_{\rm AQC}}, H_{\rm C}]||\tau \pi/4 \le \vert\vert\overline{H_{\rm AQC}}\vert\vert \tau \pi/4 $. Note that this error is accumulated at every DD application and hence if $N_D$ is the number of DD applications, then we require that $N_D \vert\vert\overline{H_{\rm AQC}}\vert\vert \tau \pi/4 < \epsilon$, for some small error, $\epsilon$.  The fidelity of the decoupling pulse may then be approximated as $\mathcal{F}\simeq 1 - \order{N_D^2 \tau^2 \vert\vert\overline{H_{\rm AQC}}\vert\vert^2}$.  The easiest way to keep the error small (and the fidelity approaching unity) is by performing the DD pulses as quickly as possible, decreasing $\tau$.  Though reducing the strength of the adiabatic Hamiltonian would also reduce the error associated with the DD, it would increase the length of the computation \emph{and} increase the susceptibility to noise, so it is inadvisable.

\subsection*{The \texorpdfstring{$\cbb{6k,2k,2}$}{[[6k,2k,2]]} code with unmatched interactions}
\noindent  The various potential physical implementations of AQC will each find certain interactions to be easier to implement than others.  Charge qubit implementations, for example, are significantly more amenable to unmatched, $XZ$- and $ZX$-type interactions than to the matched $XX,ZZ$ interactions native to the $\cbb{6k,2k,2}$ CSS code \cite{hollenberg2004charge}. However, if the interaction graph of the unencoded Hamiltonian is bipartite, then a simple Hadamard transformation on one of the bipartite halves will interconvert between matched and unmatched interaction types.  The $\cbb{6k,2k,2}$ encoding preserves the bipartite nature of the coupling graph, and so the same technique may be used after encoding to produce a new, non-CSS code whose logical operators and stabilizers have unmatched interactions.

\subsection*{Limitations of using the \texorpdfstring{$\cbb{6k,2k,2}$}{[[6k,2k,2]]} codes with AQC}
Physical noise processes that insert many-qubit operators during adiabatic computations will likely not be suppressed given that the $\cbb{6k,2k,2}$ code has a distance of $d=2$.  In most systems, though, these are far less likely events than single-qubit error operators.  Furthermore, designing codes to detect weight-two errors will require code distance of $d \geq 3$.  Unfortunately, such code will, by necessity, require logical operators of at least weight-three and thereby make the encoded Hamiltonians quite impossible to implement.

The $\cbb{6k,2k,2}$ family of codes necessitates the effective application of two weight-$4k$ DD pulses to increase the energy penalty of states not in the codespace.  However, as the problem size of the AQC increases, the amount of error suppression will reach a limit dictated by the rate, $\frac{1}{\tau}$, of applying the control Hamiltonian, $H_{\rm C}$.  This restriction is important to note but is solely due to the hardware constraints and not by the ability to encode more logical qubits in the $\cbb{6k,2k,2}$ code.

\section{Conclusion}
\label{sec:conclude}
\noindent 
We have proposed a new family of $\cbb{6k,2k,2}$ CSS codes that may be used for practical quantum error suppression in adiabatic quantum computation.  Our encoding supports a computationally universal set of matched or unmatched, weight-two interactions and preserves fixed degree and planarity properties of the original unencoded adiabatic problem upon encoding.   These codes are the first known to posses these properties, and as such, take an important step towards the physical implementation of error-suppressed adiabatic quantum computation.

\vspace{0.2in}
\begin{acknowledgments}
We express our gratitude to the AQUARIUS Architecture team at Sandia National Laboratories for their discussions on the feasibility of the $\cbb{6k, 2k, 2}$ family of error-detecting codes. In particular, we would like to thank Mohan Sarovar and Robin Blume-Kohout for illuminating discussions of error suppression techniques.  This work was supported by the Laboratory Directed Research and Development program at Sandia National Laboratories. Sandia National Laboratories is a multi-program laboratory managed and operated by Sandia Corporation, a wholly owned subsidiary of Lockheed Martin Corporation, for the U.S. Department of Energy's National Nuclear Security Administration under contract DE-AC04-94AL85000.
\end{acknowledgments}

%
%


\end{document}